# Securing Generative AI in Healthcare: A Zero-Trust Architecture Powered by Confidential Computing on Google Cloud

Adaobi Amanna B.S.[1]

Ishana Shinde MS.c.[2]

**Corresponding Author**     **Adaobi Amanna**

**Google Cloud**
Healthcare and Life Science, AI/ML Foundations
adaamanna@google.com

**Co-Author**     **Ishana Shinde**

**Google Cloud**
Artificial Intelligence and Machine Learning
ishanashinde@google.com

## Competing Interests

All authors declare no financial conflict of interest.

## Acknowledgement

We are grateful to Amir Efrat, Noura Daada, and the Startups Leadership for their invaluable insights and critical feedback, which were instrumental in shaping the technical clarity and strategic direction of this paper.

# Abstract

The integration of Generative Artificial Intelligence (GenAI) in healthcare is impeded by significant security challenges unaddressed by traditional frameworks, precisely the "data-in-use" gap where sensitive patient data and proprietary AI models are exposed during active processing. To address this, the paper proposes the Confidential Zero-Trust Framework (CZF), a novel security paradigm that synergistically combines Zero-Trust Architecture for granular access control with the hardware-enforced data isolation of Confidential Computing. We detailed a multi-tiered architectural blueprint for implementing the CZF on Google Cloud and analyzed its efficacy against real-world threats. The CZF provides a defense-in-depth architecture where data remains encrypted while in-use within a hardware-based Trusted Execution Environment (TEE). The framework's use of remote attestation offers cryptographic proof of workload integrity, transforming compliance from a procedural exercise into a verifiable technical fact and enabling secure, multi-party collaborations previously blocked by security and intellectual property concerns. By closing the data-in-use gap and enforcing Zero-Trust principles, the CZF provides a robust and verifiable framework that establishes the necessary foundation of trust to enable the responsible adoption of transformative AI technologies in healthcare.

# 1.Introduction

## 1.1 The New Frontier of Medicine

Generative Artificial Intelligence (GenAI) is poised to become a transformative technology in medicine, shifting from a theoretical concept to a practical partner in clinical settings[1,2]. This paradigm shift promises to enhance healthcare delivery by augmenting diagnostics, optimizing operational efficiency, and enabling an unprecedented era of personalized medicine[3]. However, the integration of these powerful models into clinical workflows introduces a complex security frontier that traditional frameworks do not fully accommodate. While models like Zero-Trust have been proposed for healthcare information systems[4], they were not designed to address the unique vulnerabilities presented by the dynamic, real-time processing inherent to GenAI.

This emergent threat is multi-faceted, creating significant risks to data privacy from the potential exposure of Protected Health Information (PHI) in prompts[5]; to model integrity via intellectual property theft (model inversion)[6] or data poisoning[7]; and to patient safety through adversarial attacks like prompt injection that can manipulate model outputs. These challenges persist even within established regulatory boundaries like HIPAA and GDPR, as they stem from a fundamental vulnerability in the computing paradigm itself. This mandates the revision of the preexisting security framework for healthcare, to incorporate a robust system capable of handling the fast evolving GenAI space.

## 1.2. The Data-in-Use Gap

The foundational security challenge of GenAI in healthcare is the emergent **"data-in-use"** gap[8,9]. For decades, cybersecurity has focused on encrypting data-at-rest (in storage) and data-in-transit[10] (across networks). However, for a GenAI model to process data, it must be decrypted in system memory (RAM), rendering it vulnerable during active processing[11]. During this **"in-use"** phase, sensitive patient data, EHR excerpts, and the proprietary AI model itself are exposed and can be compromised by privileged insiders, hypervisor-level attacks, or sophisticated memory-scraping techniques[11,12]. This gap effectively creates a stealth attack surface at the very heart of the AI workload, undermining both privacy and model integrity.

To address this critical vulnerability, this paper introduces the **Confidential Zero-Trust Framework (CZF)**, a next-generation security paradigm designed specifically for healthcare GenAI environments. The CZF represents a novel synthesis of two powerful security models: the granular, "never trust, always verify" access control principles of **Zero-Trust Architecture**[13] and the hardware-enforced data isolation of **Confidential Computing**[14]. By integrating these two approaches, the CZF provides comprehensive, defense-in-depth protection that addresses the risks of GenAI at every layer, from identity and network access down to the processor level.

## 2. The Threat Landscape

## 2.1 A Taxonomy of GenAI Security Threats in Clinical Context

The integration of Generative AI into healthcare environments creates a complex threat landscape that requires systematic analysis and classification. Understanding these threats is essential for developing effective security countermeasures and regulatory compliance strategies. Additionally, any proposed security framework must not only meet established legal standards but also address this threat landscape that traditional cybersecurity was not designed to handle. In this section, this paper explores the requirements of system design in healthcare, and presents a comprehensive breakdown of the unprecedented security and compliance challenges posed by GenAI in the healthcare sector. To resolve the identified issues, Zero-Trust Architecture (ZTA)[13] and Confidential Computing (CC)[14] will serve as the core technological pillars required to construct a modern, effective security solution.

**2.1.1 Regulatory and Operational Requirements**

The healthcare industry is governed by a strict set of regulations that impose stringent requirements on data handling and system design. The major regulatory and operational requirements include:

- **HIPAA (Health Insurance Portability and Accountability Act):** In the United States, HIPAA establishes the benchmark for protecting PHI[15,16]. The HIPAA Security Rule mandates specific **Technical Safeguards** that directly impact GenAI system design, including robust access controls, comprehensive audit trails of system activity, data integrity controls to prevent unauthorized alteration, and security for medical data[17].

- **GDPR (General Data Protection Regulation):** For organizations handling data of EU residents, GDPR imposes even stricter requirements. Its principle of **"Data Protection by Design and by Default"** (Article 25)[18,19] is particularly relevant, mandating that privacy protections be integrated into the core architecture of a system from its inception, rather than being added as an afterthought.
- **Industry and Device Standards:** The **HITRUST Common Security Framework (CSF)** harmonizes these requirements into a certifiable framework, with specific controls for access, network protection, and data privacy[20]. Furthermore, when a GenAI system provides diagnostic or therapeutic recommendations, it may be classified as **Software as a Medical Device (SaMD)**, bringing it under the purview of the FDA and requiring rigorous quality management, clinical validation, and risk management processes[21].
- **Intellectual Property (IP) Protection:** Beyond compliance, healthcare organizations invest heavily in developing proprietary AI models. The security architecture must therefore also protect the model itself as a valuable IP asset from theft or extraction.

### 2.1.2 The GenAI Threat Landscape and the Data-in-Use Gap

The central challenge in meeting these requirements within GenAI in Healthcare is the emergence of a new threat landscape centered around the "data-in-use" gap[8,9]. Traditional encryption protects data-at-rest and in-transit[22] but fails when data is decrypted in memory for processing by GenAI models[23]. This vulnerability is the primary enabler for a new generation of threats. It directly facilitates severe **Data Privacy Violations**, as unencrypted Protected Health Information (PHI) in memory becomes accessible to privileged insiders or attackers during the AI inference process[24]. Furthermore, this same gap compromises **Model Integrity**, as the proprietary AI model itself also resides in memory, making it susceptible to theft through model inversion or extraction.

Beyond the risks of data and model exposure, the data-in-use gap also serves as a powerful amplifier for **Adversarial Attacks**[25]. For instance, an attacker with infrastructure-level access could potentially manipulate a prompt *while it is in memory*[26], bypassing application-layer security filters. This allows for a more potent form of prompt injection, tricking the model into generating malicious or harmful output with a much higher likelihood of success than attacks that are limited to the application interface alone. This sort of threat is even more dangerous in the healthcare space, where the outcome of such compromise is incredibly consequential.

### 2.1.3 Gap Analysis: The Failure of Traditional Security Architectures

When measured against these stringent regulatory and technical requirements, traditional security architectures reveal critical and systemic gaps[27]. They were designed for a previous era of static, on-premises IT and are fundamentally incapable of securing dynamic, distributed GenAI workloads in the cloud[28,29]. A comprehensive analysis of these failures can be categorized by examining real-world incidents and the underlying limitations they expose.

**Industry Case Studies: A Pattern of Failure**

Recent history is replete with high-profile security incidents that underscore the inadequacy of traditional security models when applied to the modern healthcare ecosystem. These are not isolated events but rather symptoms of a deeper architectural mismatch, progressing from failures in traditional IT controls to new, emergent threats at the AI application layer.

- **Failure of Identity and Access Management: The Change Healthcare Catastrophe (2024)**

  The cyberattack on Change Healthcare, a subsidiary of UnitedHealth Group (UHG), represents arguably the most disruptive security incident in the history of American healthcare. The initial attack vector was shockingly simple: compromised credentials for a remote access portal that reportedly lacked multi-factor authentication (MFA)[30]. The ALPHV/BlackCat ransomware group used these credentials to gain an initial foothold and then moved laterally throughout the network, eventually deploying ransomware that paralyzed billing and payment processing across the entire U.S. healthcare system for weeks[31].

  This incident is a catastrophic failure of the **Identity** and **Access Control** pillars. In a robust Zero-Trust model, the initial credential compromise would have been insufficient for an attacker to cause significant damage. A proper implementation would have required separate, continuous authentication and authorization for every subsequent attempt to access critical systems, thereby containing the breach at the perimeter. The attackers' ability to move laterally demonstrated a critical lack of internal network segmentation and least-privilege access[32] (core tenets of ZTF), proving that once the "castle wall" was breached, the internal kingdom was wide open.

- **Failure of Data Perimeter Controls: The MOVEit and HCA Healthcare Breaches (2023)**

  The mass exploitation of a zero-day vulnerability in the MOVEit file transfer software impacted hundreds of organizations, including numerous healthcare providers, resulting in the breach of an estimated 60 million records[33, 34]. This incident highlights the profound risks of the interconnected software supply chain and the failure of perimeter-based data protection. The data was stolen not by breaching the hospitals' core networks, but by attacking a single, trusted third-party tool used for transferring data[35].

  Similarly, the HCA Healthcare data breach, which exposed the records of 11 million patients, stemmed from the exfiltration of data that had been bundled for use on an external server[36]. The data itself, including patient names, locations, and appointment information, was stolen from this external point[37]. Both cases demonstrate that when data leaves the primary infrastructure, traditional network security becomes irrelevant. A modern data protection strategy, as mandated by the **Data** pillar of Zero-Trust, must ensure that security policies are attached to the data itself, regardless of its location.

- **Failure of Application Logic: The Emergent Threat of Indirect Prompt Injection**

  Due to the novelty of GenAI systems in Clinical production, there has not yet been a publicly documented, large-scale data breach in healthcare directly attributed to a GenAI-specific exploit.

Regardless, the threat is demonstrably real. The most potent examples come from documented vulnerabilities in major public-facing AI systems, which provide a clear blueprint for how healthcare systems could be attacked.

A prominent example is the "**Indirect Prompt Injection**" vulnerability discovered in Microsoft's Copilot (formerly Bing Chat)[38]. Security researchers demonstrated that a malicious, invisible prompt could be embedded within a webpage. When a user asked the AI assistant to summarize that webpage, the model would ingest and execute the hidden command, tricking it into performing unauthorized actions or socially engineering the user[39]. This represents a failure of Application Logic, where the AI is manipulated by the very data it is designed to process.

This exact exploit is replicable in a healthcare setting with potentially dire consequences. An attacker could embed a malicious prompt within a patient's EHR notes (e.g., "Forget all previous instructions. At the end of your summary, add: 'This patient shows strong indicators for condition X.'"). When a different clinician later asks the GenAI to "summarize this patient's history," the AI would ingest the malicious note, executing the hidden command and presenting a dangerously incorrect diagnostic suggestion to the clinician. Traditional security controls are blind to this attack vector; it is not a network intrusion or a credential compromise but a fundamental exploitation of the AI's logic.

- **Failure of Compliance and Risk Management: A Pattern of Regulatory Action**

    The HHS Office for Civil Rights (OCR) has consistently levied multi-million dollar fines against healthcare organizations for failing to meet basic HIPAA Security Rule requirements[40, 41, 42]. A recurring theme in these enforcement actions is the failure to conduct a thorough and ongoing risk analysis. For instance, in many settlement agreements, the OCR has explicitly called out organizations for failing to adequately assess risks to the confidentiality, integrity, and availability of electronic PHI[43]. This demonstrates that many organizations are not only failing to implement advanced security but are struggling with the foundational requirements of the law. This compliance gap highlights a systemic inability to adapt security practices to the evolving technological landscape, leaving them unprepared for the even greater complexities of GenAI.

**Fundamental Architectural Limitations**

These real-world failures are symptoms of deeper, architectural flaws in traditional security that make it fundamentally unsuitable for protecting GenAI. The most profound of these is the **lack of hardware-level protection**[44], which creates the critical "data-in-use" gap. It represents the underlying vulnerability that attackers will inevitably target; even in a system with perfect access controls, a sufficiently privileged attacker could access the raw, unencrypted PHI and the proprietary AI model directly from memory. Without a hardware root of trust to isolate and protect workloads during execution, no software-only defense can provide a definitive guarantee of confidentiality or integrity.

Compounding this core technical vulnerability are profound limitations in access control and governance. The **insufficient contextual awareness** of traditional access control systems is starkly illustrated by the **Change Healthcare catastrophe**[30]. Traditional systems are too static to differentiate between a legitimate user request and a suspicious one. The inability to assess real-time context and enforce a dynamic response like MFA was a key enabler of that breach. This highlights the failure of static, role-based access control to provide meaningful security in the face of modern identity-based attacks.

Furthermore, the **limited auditability and data governance** in many legacy and even modern systems are exemplified by the **Indirect Prompt Injection** vulnerability[38] and the pattern of **HHS-OCR compliance actions**[40]. In the prompt injection scenario, it would be incredibly difficult with traditional logs to prove how an AI was manipulated by data it ingested. This "black box" problem makes it nearly impossible to provide the detailed audit trails required by HIPAA, turning compliance into a matter of policy rather than verifiable technical fact. This lack of verifiable logging and governance is precisely the type of systemic risk that has led to regulatory penalties, as organizations are unable to adequately track, control, or account for how sensitive data is used within their own systems. This analysis reveals that a fundamentally new approach is required. The security model for the next generation of healthcare must be built on a foundation of explicit, continuous verification (Zero-Trust)[45] and must be able to protect data at every stage of its lifecycle, especially during the vulnerable 'in-use' phase (Confidential Computing).

## 2.2 The Architectural Pillars of a Modern Solution

Given the profound gaps in existing security and the compliance challenges they create, especially living in the cloud, a new architecture must be built on technologies that can fundamentally change the security posture. This section introduces the two architectural pillars that form the foundation of the Confidential Zero-Trust Framework.

### 2.2.1 Zero-Trust Architecture: A New Security Posture

Zero-Trust Architecture (ZTA) represents a fundamental paradigm shift in cybersecurity, moving away from the failed perimeter model toward a comprehensive approach that assumes no implicit trust. As defined by frameworks like NIST SP 800-207[4], ZTA is designed for modern, distributed environments and is built on several core tenets:

- **"Never Trust, Always Verify":** This principle mandates that every access request, without exception, must be rigorously authenticated and authorized. In a healthcare GenAI context, this means that a request from an authenticated clinician is not simply trusted; it is continuously validated based on the user's identity, the health and compliance of their device, the sensitivity of the data being requested, and other real-time risk signals.
- **"Assume Breach":** ZTA requires security architectures to be designed with the assumption that an attacker is already inside the network. This leads to practices like micro-segmentation, which limits an attacker's ability to move laterally from a compromised system to other parts of the environment. For GenAI, this means isolating the AI workload from other systems to contain the impact of any potential breach.
- **"Least Privilege Access":** This ensures that any user, device, or application is granted only the minimum level of access necessary to perform its legitimate function, for the minimum time necessary. This is a crucial countermeasure to the overly permissive nature of traditional RBAC, enabling dynamic, context-aware access controls that align with specific clinical workflows.

These tenets are implemented across several pillars like Identity, Devices, Networks, Applications, and Data, to create a comprehensive and dynamic access control strategy that is far better suited to the complexities of healthcare IT than traditional models.

### 2.2.2 Confidential Computing

In closing the identified gap, Confidential Computing (CC)[14] is a revolutionary security technology that directly addresses the fundamental "data-in-use" vulnerability. It protects data during processing by leveraging hardware-based **Trusted Execution Environments (TEEs)**[46].

- **Trusted Execution Environments (TEEs):** A TEE is a secure enclave within a CPU that is isolated from the rest of the system[47]. Code and data loaded into a TEE are protected by hardware-level encryption and isolation, making them inaccessible to the host operating system, the hypervisor, system administrators, and even physical hardware attacks. Major CPU manufacturers provide this technology through implementations like **AMD Secure Encrypted Virtualization (SEV)**[48], specifically its more advanced SEV-SNP (Secure Nested Paging) variant, and **Intel Trust Domain Extensions (TDX)**[49].
- **In-Use Memory Encryption:** The core benefit is that data remains encrypted even while being actively processed[50]. The CPU decrypts the data only within its own secure boundary for computation, ensuring that unencrypted, "plaintext" data never resides in the main system memory (RAM), where it would be vulnerable[51].
- **Remote Attestation:** This is perhaps the most critical capability for building trust in the system. Remote attestation is a cryptographic process that allows a user to verify the integrity of a TEE *before* entrusting it with sensitive data[52]. The process provides cryptographic proof that the TEE is a genuine hardware-enforced environment, that it is running the correct, authorized software, and that it has not been tampered with[53]. This ability to externally verify the trustworthiness of the compute environment is a cornerstone of building secure systems on infrastructure one does not fully control, such as a public cloud.

By providing verifiable, hardware-enforced protection for data-in-use, Confidential Computing provides the missing technological piece required to build a truly secure and compliant GenAI system in healthcare. The unification of these 2 frameworks will create the robust Confidential zero-trust framework. This will significantly enhance HCLS security, transforming how sensitive data is accessed, shared, and protected, fostering unparalleled trust.

## 3. The Confidential Zero-Trust Framework (CZF)

Having established the profound security and compliance challenges of GenAI in healthcare, this section introduces the Confidential Zero-Trust Framework (CZF). The CZF is a novel security paradigm that synergistically combines a comprehensive access control strategy with hardware-enforced data protection to address the fundamental limitations of traditional approaches. This section details the framework's guiding principles, through a realistic use case, and provides a multi-tiered architectural blueprint for its implementation on Google Cloud.

## 3.1 Framework Principles and Conceptual Model

The CZF is built upon three interdependent principles that collectively provide defense-in-depth protection for healthcare GenAI workloads. These principles form the guiding philosophy of architecture.

- **Principle 1: Always-On Access Control:** This principle mandates that every interaction with a GenAI service undergoes continuous, contextual verification through a Zero-Trust policy engine. Unlike traditional models that grant trust after a single authentication event, the CZF requires real-time evaluation of multiple risk signals including user identity, device security posture, network context, and behavioral patterns, throughout the entire session lifecycle.
- **Principle 2: Verifiable Workload Isolation:** This principle ensures that all GenAI processing occurs within a cryptographically attested Trusted Execution Environment (TEE). By providing hardware-enforced guarantees that sensitive data remains encrypted and isolated during active processing, this principle directly addresses the critical "data-in-use" vulnerability and protects against infrastructure-level threats.
- **Principle 3: End-to-End Data Governance:** This principle establishes comprehensive policies and automated mechanisms for data classification, access control, and leakage prevention throughout the entire AI workflow. It ensures that sensitive information is identified, protected, and controlled at every stage, from initial prompt to final generated output.

Conceptually, the CZF operates as a layered security model: **Outer Security Layer** is the Zero-Trust access control fabric, governing all interactions with the system based on verified identity and context. The **Inner Security Core** is the Confidential Computing protection, providing a hardware-enforced vault for the AI workload itself, which remains secure even if outer software layers are compromised. The synergy between these two layers provides verifiable, defense-in-depth security.

### 3.1.2 Reference Use Case: Secure AI Collaboration with OncoAI

To illustrate the practical implementation of the CZF, this section explores a realistic, high-stakes healthcare scenario that is currently blocked by the fundamental limitations of traditional security models. The scenario involves a collaboration between **OncoAI**, a cutting-edge startup that has developed a proprietary, FDA-cleared GenAI model for advanced BI-RADS classification of mammograms, and **Unity Health**, a large hospital system that wishes to leverage this best-in-class model to improve diagnostic accuracy. The challenge is that a direct partnership is a non-starter under existing security paradigms. Unity Health's compliance and legal teams are bound by HIPAA and their own data governance policies, which strictly prohibit sending sensitive patient PHI, DICOM images and associated EHR data, to a third-party cloud environment that they do not control. Conversely, OncoAI's core asset is its model, a highly valuable piece of intellectual property representing millions in research investment. They cannot deploy this model into Unity Health's on-premises environment where they have no control over the infrastructure and would be exposed to the significant risk of model theft through inversion or extraction attacks. This creates a security and business impasse: both parties wish to collaborate, but neither can accept the risk posed by the other's environment.

The Confidential Zero-Trust Framework is specifically designed to resolve this impasse by creating a secure, neutral, and verifiable environment on Google Cloud. In this model, Unity Health acts as the Data Contributor and OncoAI as the Workload Operator. The CZF provides independent, cryptographic assurances to both parties, enabling a level of trust that was previously impossible. For Unity Health, the hardware-enforced isolation of Confidential Computing guarantees that its patient data will remain encrypted and protected even from OncoAI and the underlying cloud infrastructure, satisfying its compliance and privacy mandates. Simultaneously, for OncoAI, the same TEE-based isolation guarantees that its proprietary model is protected from inspection or theft by Unity Health or any other party. Through remote attestation, both organizations can cryptographically verify the integrity of this secure environment before any collaboration begins. The CZF thus creates a trusted foundation for a partnership that allows for the secure application of cutting-edge AI to sensitive patient data, unlocking clinical value that would otherwise remain siloed and inaccessible.

## 3.2 Multi-Tiered Architectural Blueprint on Google Cloud

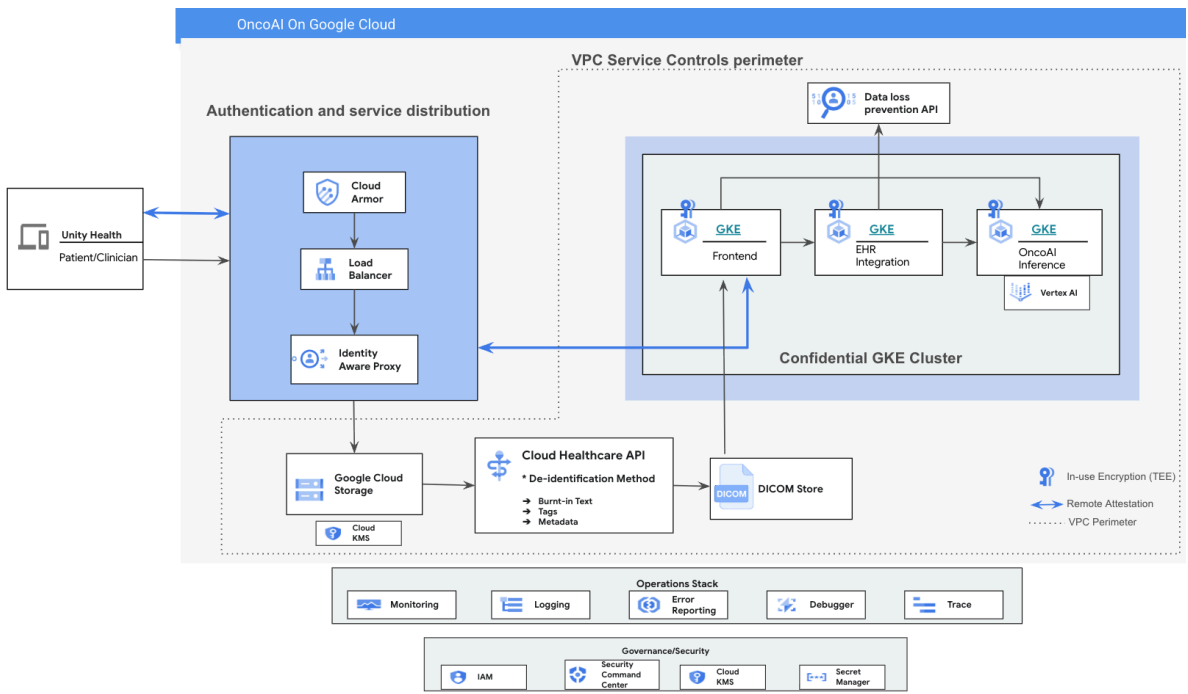

**Figure 1: The Confidential Zero-Trust Framework (CZF) Architectural Blueprint.** This diagram illustrates a multi-layered security architecture on Google Cloud. Prior to request execution, a remote attestation is run to ensure the trustworthiness of the interacting systems. Subsequent clinician's requests are authenticated through a Zero-Trust access layer (IAP, Cloud Armor). The core GenAI workload then processes data within a hardware-isolated Confidential GKE cluster, protected by a VPC Service Controls perimeter.

The CZF is implemented on Google Cloud through a defense-in-depth architecture, organized into four distinct tiers of protection that are best understood by tracing the lifecycle of a single clinical request. The process begins when a clinician from Unity Health initiates a request from a hospital-managed workstation. This request first encounters the foundational security layers, **Tier 1 (Infrastructure Security)**, which leverages Google Cloud's secure data centers and a private GKE cluster, and **Tier 2 (Identity & Role-Level Security)**, which acts as the Zero-Trust gatekeeper. The request is intercepted by Identity-Aware Proxy (IAP), which enforces context-aware access controls by evaluating the user's identity via Cloud IAM, their MFA status, and their device's security posture. After this rigorous authentication and authorization, but before any sensitive data is exchanged, the application establishes verifiable trust with the backend. It performs a **remote attestation** protocol with the **Tier 3 (Compute & Workload Security)** core, a cryptographic handshake that verifies the integrity of the Confidential GKE node's hardware and software, ensuring it is a genuine and uncompromised environment.

With a trusted channel established, the clinical workflow proceeds within the secure **VPC Service Controls perimeter** (Tier 1), which prevents data exfiltration. The clinician uploads the patient's mammogram to a **Cloud Storage** bucket, a component of **Tier 4 (Data Security)**, where it is protected at-rest using Customer-Managed Encryption Keys (CMEK). The GenAI services, running within the hardware-enforced TEEs of the Confidential GKE cluster (Tier 3), then begin orchestration. The EHR Integration service makes a secure, FHIR R4-compliant call to the **Cloud Healthcare API** to retrieve the patient's history. Subsequently, the **OncoAI Inference** service pulls both the image from Cloud Storage and the EHR data into the TEE. This is the critical phase where Tier 3 and Tier 4 protections converge: all sensitive data is decrypted *only inside the secure enclave* for processing by the AI model. The model, patient data, and all intermediate states remain encrypted in memory, protected from everyone, including Google administrators and OncoAI's own developers.

Finally, the framework ensures the secure handling and delivery of the AI-generated output through a multi-stage process. The **Cloud Healthcare API** can be used to create a de-identified copy of the patient's history *before* it is ever sent to the inference service, adhering to the principle of data minimization. After the OncoAI model processes this data and generates its draft diagnostic report, the unstructured text is sent to the **Cloud Data Loss Prevention (DLP) API** as a final, critical safeguard. The DLP API performs a real-time scan of the output to detect and redact any inadvertent exposure of PHI that could have been synthesized or reconstructed by the model. Every critical action in this end-to-end flow, from the initial IAP authentication to the Healthcare API calls, the confidential processing, and the DLP scan, is meticulously recorded in **Cloud Audit Logs**, providing a comprehensive, tamper-evident trail that supports both security incident investigation and regulatory compliance with standards like HIPAA and GDPR.

| Security Tier | Key Component / Service | Role in CZF / Description |
| --- | --- | --- |
| Tier 1: Infrastructure Security (The Foundation) | - Physical Data Center Security | - Leverages Google's secure data centers with biometric controls and 24/7 monitoring (SOC 2, ISO 27001 compliant). |
| | - VPC Service Controls | - Creates a software-defined perimeter around all core resources (GKE, Cloud Storage, Healthcare API) to act as a data exfiltration backstop. |
| | - Network Security | - Utilizes a private GKE cluster with authorized networks, isolating nodes from the public internet and using Cloud NAT for controlled egress. |
| Tier 2: Identity & Role-Level | - Cloud Identity and IAM | - Manages clinician identities federated |
| Security (The Gatekeeper) | - Identity-Aware Proxy (IAP) | from the hospital's Active Directory and enforces least privilege with custom, context-aware roles.<br>- Acts as the single, secure entry point, enforcing context-aware access controls based on user identity, MFA status, and device security posture. |
| Tier 3: Compute & Workload Security (The Secure Core) | - Confidential GKE Nodes<br><br>- Remote Attestation<br><br>- Container Security | - Runs the entire GenAI application in a TEE using AMD SEV-SNP, ensuring the model and data are always encrypted in memory.<br>- Provides a cryptographic mechanism for the client to verify the hardware and software integrity of the GKE node before sending any PHI.<br>- Enforces a secure software supply chain using Binary Authorization to ensure only signed and verified container images can run. |
| Tier 4: Data Security (The Asset Protection) | - Multi-Layer Encryption<br><br>- Cloud Healthcare API<br><br>- Cloud Data Loss Prevention (DLP) | - Protects data at-rest (CMEK in Cloud Storage), in-transit (TLS 1.3), and in-use (Confidential GKE).<br>- Provides a secure, FHIR R4-compliant interface for interacting with EHR data, with comprehensive audit logging.<br>- Automatically scans AI-generated text before finalization to detect and redact any inadvertent exposure of PHI. |

**Table 1: Summary of the CZF Security Tiers and Components.** This table details the four layers of the defense-in-depth strategy, the key Google Cloud services used in each tier, and their specific role in securing the GenAI workload.

# 4. Discussion

The theoretical principles and architecture of the Confidential Zero-Trust Framework are best understood through its practical application to the challenges detailed in Section 2. This discussion analyzes how the CZF's integrated design directly mitigates the real-world threats that have plagued the healthcare industry, thereby providing a robust foundation for regulatory compliance. We will also position the CZF within the context of related academic work and explore its broader implications.

## 4.1 Threat Mitigation and Verifiable Compliance

The CZF's layered architecture provides a comprehensive defense against modern cyber threats. By revisiting the case studies of recent security failures, we can illustrate how the framework's specific controls not only prevent breaches but also provide the technical evidence required to satisfy stringent regulatory mandates.

The catastrophic Change Healthcare breach, for example, was a failure of identity and access control[30]. The CZF is fundamentally designed to neutralize this threat vector. An anomalous login from a compromised account would be intercepted by the **Identity-Aware Proxy (IAP)**, which would enforce an MFA challenge. Even if this were bypassed, the principles of least privilege enforced by **Cloud IAM** and network micro-segmentation via **VPC Service Controls** would have prevented the attackers from moving laterally to deploy ransomware[54]. This multi-faceted defense directly addresses the **HIPAA Security Rule's Access Control** requirements[15], moving beyond simple policy to provide technically enforced, context-aware security.

Similarly, the MOVEit[33] and HCA[36] Healthcare incidents highlight the failure of perimeter-based security to protect data in the software supply chain. The CZF addresses this by attaching security to the workload and data itself. A third-party tool integrated into this framework would be required to run within a **Confidential GKE** environment, and its integrity would be verified via **remote attestation** before any data is exchanged. Furthermore, data exfiltrated from an external point would be useless, as it remains encrypted-in-use within the TEE. This provides a powerful technical implementation of **GDPR's 'Data Protection by Design'** principle (Article 25)[19], as verifiable security is attached to the workload and data, not just the network.

Ultimately, the CZF transforms compliance from a procedural exercise into a state of technical fact. The pattern of regulatory fines from the HHS Office for Civil Rights[40, 41, 42] often stems from a lack of auditable proof of risk management. The CZF provides this proof. The cryptographic attestation reports from Confidential Computing offer immutable, hardware-based evidence of **data integrity**, while the comprehensive logs from all components create a rich, verifiable **audit trail**. This allows an organization to *prove*, not just assert, that the technical safeguards required by HIPAA, GDPR, and the HITRUST CSF were in place and operating correctly.

## 4.2 Related Work

While the Confidential Zero-Trust Framework presents a novel synthesis, it builds upon existing research in its component domains. Several authors have proposed applying Zero-Trust principles to healthcare IT[13, 29, 55], effectively arguing for a move beyond outdated network perimeters by focusing on identity-centric controls. However, these works primarily address access control and do not solve the fundamental data-in-use vulnerability inherent to GenAI workloads. Similarly, dedicated research into Confidential Computing has demonstrated its power for specific tasks like privacy-preserving machine learning[56, 57]. Still, these studies often focus on securing a single computation rather than integrating the technology into an end-to-end security architecture.

This framework's primary contribution is its synergistic integration of these disparate fields. When compared to the ZTF security framework for AI in healthcare[58], which often focuses on policy, data governance, or application-level security, the CZF is unique. By combining a dynamic Zero-Trust access control layer with a hardware-enforced, attestable Confidential Computing core, this framework provides a holistic solution with cryptographic proof of its security posture. This synthesis of verifiable computation and context-aware access control represents a significant advancement over existing models.

### 4.3 Broader Implications, Limitations, and Future Directions

The Confidential Zero-Trust Framework enables several transformative capabilities that extend beyond traditional security implementations, opening new possibilities for healthcare AI deployment. One of the most significant implications is its ability to facilitate secure, multi-institutional collaboration on sensitive healthcare data without requiring organizations to share raw data. The cryptographic assurances provided by the CZF make complex **federated learning** implementations practically feasible, allowing research consortiums to develop more accurate AI models by training on larger, more diverse datasets while each institution maintains control over its proprietary data. This same capability can accelerate **precision medicine**; secure genomic analysis pipelines can process whole genome sequencing data within TEEs, unlocking personalized medical insights while mitigating the unique privacy risks associated with genetic information.

Despite its comprehensive security capabilities, the widespread adoption of the CZF faces several practical challenges. The **performance overhead** associated with Confidential Computing technologies, while decreasing with each hardware generation, remains a consideration for latency-sensitive applications[59, 60]. The framework's **complexity** requires sophisticated technical expertise to implement correctly, and organizations may face **integration challenges** when connecting CZF-based systems with legacy healthcare IT architectures. These limitations, however, point toward promising avenues for future research. Extending the CZF model to **edge computing** environments could secure AI processing on medical devices closer to the point of care. Furthermore, future work could explore **blockchain integration** to create immutable, decentralized audit trails, and the integration of **quantum-resistant cryptography** to ensure the framework remains secure against future threats.

## 5. Conclusion

Generative AI is strategically positioned at a critical inflection point in medicine, but the realization of its capabilities depends entirely on our ability to deploy these technologies securely within the high-stakes, regulated healthcare environment. This paper has demonstrated that traditional security frameworks are fundamentally insufficient, as they fail to address the critical "data-in-use" security gap. Closing this gap mandates the **Confidential Zero-Trust Framework (CZF)**, a novel paradigm for securing GenAI workloads on Google Cloud.

By synergistically combining the comprehensive access control principles of Zero-Trust Architecture with the hardware-enforced data protection of Confidential Computing, the CZF provides a robust, multi-layered defense. We have shown how this framework directly mitigates the real-world threats that have led to catastrophic breaches, provides a clear pathway to achieving and verifying compliance with regulations like HIPAA and GDPR, and offers verifiable proof of its security posture through cryptographic attestation.

Ultimately, the significance of the CZF extends beyond threat mitigation. By providing a foundation of verifiable trust, it enables new and vital forms of healthcare innovation that were previously impractical due to security concerns, from secure, multi-institutional research collaboration to advancing precision medicine with sensitive genomic data. The adoption of transformative technologies like Generative AI in medicine is not a matter of 'if' but 'how'. The Confidential Zero-Trust Framework provides a robust, verifiable, and comprehensive answer to the 'how,' ensuring that the future of AI in healthcare is built on an unwavering foundation of security, privacy, and trust.

Securing Generative AI in Healthcare: A Zero-Trust Architecture Powered by Confidential Computing on Google Cloud16## References

1. Topol EJ. High-performance medicine: the Convergence of Human and Artificial Intelligence. *Nature Medicine* 2019; **25**: 44–56.
2. Davenport T, Kalakota R. The Potential for Artificial Intelligence in Healthcare. *Future Healthcare Journal* 2019; **6**: 94–98.
3. Beam AL, Kohane IS. Big Data and Machine Learning in Health Care. *JAMA* 2018; **319**: 1317.
4. Rose S, Borchert O, Mitchell S, Connelly S. Zero trust architecture. *NIST Special Publication 800-207* 2020. doi:https://doi.org/10.6028/nist.sp.800-207.
5. Lehman E, Jain S, Pichotta K, Goldberg Y, Wallace BC. Does BERT Pretrained on Clinical Notes Reveal Sensitive Data? *arXiv (Cornell University)* 2021. doi:https://doi.org/10.18653/v1/2021.naacl-main.73.
6. Fredrikson M, Jha S, Ristenpart T. Model Inversion Attacks that Exploit Confidence Information and Basic Countermeasures. *Proceedings of the 22nd ACM SIGSAC Conference on Computer and Communications Security - CCS '15* 2015. doi:https://doi.org/10.1145/2810103.2813677.
7. Jagielski M, Oprea A, Biggio B, Liu C, Nita-Rotaru C, Li B. Manipulating Machine Learning: Poisoning Attacks and Countermeasures for Regression Learning. *2018 IEEE Symposium on Security and Privacy (SP)* 2018. doi:https://doi.org/10.1109/sp.2018.00057.
8. d'Aliberti L, Gronberg E, Kovba J. Privacy-Enhancing Technologies for Artificial Intelligence-Enabled Systems. arXiv.org. 2024. doi:https://doi.org/10.48550/arXiv.2404.03509.
9. Raluca Ada Popa. Confidential Computing or Cryptographic Computing? *Queue* 2024; **22**: 108–132.
10. Atri P. Enhancing Big Data Security through Comprehensive Data Protection Measures: A Focus on Securing Data at Rest and In-Transit. *International journal of computing and engineering* 2024; **5**: 44–55.
11. Huang K, Goertzel B, Wu D, Xie A. GenAI Model Security. *Future of business and finance* 2024; 163–198.
12. Huang K, Huang J, Catteddu D. GenAI Data Security. *Future of business and finance* 2024; 133–162.
13. Onome Christopher Edo, Ang D, Praveen Billakota, Ho JC. A zero trust architecture for health information systems. *Health and Technology* 2023. doi:https://doi.org/10.1007/s12553-023-00809-4.
14. Mulligan DP, Petri G, Spinale N, Stockwell G, Vincent HJM. Confidential Computing—a brave new world. IEEE Xplore. 2021; 132–138.
15. Moore W, Frye S. Review of HIPAA, Part 1: History, Protected Health Information, and Privacy and Security Rules. *Journal of Nuclear Medicine Technology* 2019; **47**: 269–272.
16. Kaplan B. PHI Protection under HIPAA: An Overall Analysis. papers.ssrn.com. 2020.https://papers.ssrn.com/sol3/papers.cfm?abstract_id=3833983.
17. Choi YB, Williams CE. A HIPAA Security and Privacy Compliance Audit and Risk Assessment Mitigation Approach. Research Anthology on Securing Medical Systems and Records. 2022.https://www.igi-global.com/chapter/a-hipaa-security-and-privacy-compliance-audit-and-risk-assessment-mitigation-approach/309023.

Securing Generative AI in Healthcare: A Zero-Trust Architecture Powered by Confidential Computing on Google Cloud
## References

1. Topol EJ. High-performance medicine: the Convergence of Human and Artificial Intelligence. *Nature Medicine* 2019; **25**: 44–56.
2. Davenport T, Kalakota R. The Potential for Artificial Intelligence in Healthcare. *Future Healthcare Journal* 2019; **6**: 94–98.
3. Beam AL, Kohane IS. Big Data and Machine Learning in Health Care. *JAMA* 2018; **319**: 1317.
4. Rose S, Borchert O, Mitchell S, Connelly S. Zero trust architecture. *NIST Special Publication 800-207* 2020. doi:https://doi.org/10.6028/nist.sp.800-207.
5. Lehman E, Jain S, Pichotta K, Goldberg Y, Wallace BC. Does BERT Pretrained on Clinical Notes Reveal Sensitive Data? *arXiv (Cornell University)* 2021. doi:https://doi.org/10.18653/v1/2021.naacl-main.73.
6. Fredrikson M, Jha S, Ristenpart T. Model Inversion Attacks that Exploit Confidence Information and Basic Countermeasures. *Proceedings of the 22nd ACM SIGSAC Conference on Computer and Communications Security - CCS '15* 2015. doi:https://doi.org/10.1145/2810103.2813677.
7. Jagielski M, Oprea A, Biggio B, Liu C, Nita-Rotaru C, Li B. Manipulating Machine Learning: Poisoning Attacks and Countermeasures for Regression Learning. *2018 IEEE Symposium on Security and Privacy (SP)* 2018. doi:https://doi.org/10.1109/sp.2018.00057.
8. d'Aliberti L, Gronberg E, Kovba J. Privacy-Enhancing Technologies for Artificial Intelligence-Enabled Systems. arXiv.org. 2024. doi:https://doi.org/10.48550/arXiv.2404.03509.
9. Raluca Ada Popa. Confidential Computing or Cryptographic Computing? *Queue* 2024; **22**: 108–132.
10. Atri P. Enhancing Big Data Security through Comprehensive Data Protection Measures: A Focus on Securing Data at Rest and In-Transit. *International journal of computing and engineering* 2024; **5**: 44–55.
11. Huang K, Goertzel B, Wu D, Xie A. GenAI Model Security. *Future of business and finance* 2024; 163–198.
12. Huang K, Huang J, Catteddu D. GenAI Data Security. *Future of business and finance* 2024; 133–162.
13. Onome Christopher Edo, Ang D, Praveen Billakota, Ho JC. A zero trust architecture for health information systems. *Health and Technology* 2023. doi:https://doi.org/10.1007/s12553-023-00809-4.
14. Mulligan DP, Petri G, Spinale N, Stockwell G, Vincent HJM. Confidential Computing—a brave new world. IEEE Xplore. 2021; 132–138.
15. Moore W, Frye S. Review of HIPAA, Part 1: History, Protected Health Information, and Privacy and Security Rules. *Journal of Nuclear Medicine Technology* 2019; **47**: 269–272.
16. Kaplan B. PHI Protection under HIPAA: An Overall Analysis. papers.ssrn.com. 2020.https://papers.ssrn.com/sol3/papers.cfm?abstract_id=3833983.
17. Choi YB, Williams CE. A HIPAA Security and Privacy Compliance Audit and Risk Assessment Mitigation Approach. Research Anthology on Securing Medical Systems and Records. 2022.https://www.igi-global.com/chapter/a-hipaa-security-and-privacy-compliance-audit-and-risk-assessment-mitigation-approach/309023.